\documentclass[10pt]{article}

\usepackage{doublespace}
\usepackage{note}
\usepackage{epsfig}
\usepackage{amssymb}

\renewcommand{\be}{\begin{equation}}
\renewcommand{\ee}{\end{equation}}
\renewcommand{\bea}{\begin{eqnarray}}
\renewcommand{\eea}{\end{eqnarray}}
\renewcommand{\eq}[1]{Eq.~(\ref{eq:#1})}
\renewcommand{\eqs}[2]{Eqs.~(\ref{eq:#1}) and (\ref{eq:#2})}
\newcommand{\B}[1]{|\Phi_#1\>}
\newcommand{\refcite}[1]{\cite{#1}}
\def\<{\langle}
\def\>{\rangle}

\def\lpm{ \left(\rule{0pt}{2.1ex}\right. }
\def\rpm{ \left.\rule{0pt}{2.1ex}\right) }

\def\dag{\dagger}
\def\ot{\otimes}

\def\non{\nonumber}
\def\os2{{1 \over \sqrt{2}}}
\def\s{\sigma}

\def\RR{\mathbb R}
\begin{document}

\markboth{Debbie W. Leung}
{Quantum computation by measurements}


\title{Quantum computation by measurements}

\author{Debbie W. Leung}
\address{Institute of Quantum Information, MSC 107-81, Caltech, 
Pasadena, CA 91125, USA \\
wcleung\@@caltech.edu}

\date{October 31, 2003}


\maketitle


\begin{abstract}
We first consider various methods for the indirect implementation 
of unitary gates.  
We apply these methods to rederive the universality of $4$-qubit
measurements based on a scheme much simpler than Nielsen's original
construction [quant-ph/0108020].
Then, we prove the universality of simple discrete sets of $2$-qubit
measurements, again using a scheme simplifying the initial
construction [quant-ph/0111122].
Finally, we show how to use a single $4$-qubit measurement to achieve
universal quantum computation, and outline a proof for the
universality of almost all maximally entangling $4$-qubit
measurements.
\end{abstract}


\section{Introduction}

Studying the resources required for universal quantum computation is
important not only for its realization but also for our theoretical
understanding of what makes quantum computation so powerful.  
The standard model of quantum computation\cite{DiVincenzo95a}
requires a well defined and isolated Hilbert space.
Universal computation further requires the ability (1) to prepare
a fiducial initial state, 
(2) to implement a {\em universal} set of gates in a quantum circuit, 
and (3) to perform strong measurements.  
A set of quantum gates is universal if any unitary evolution can
be approximated to arbitrary accuracy by a circuit involving those
gates only.\cite{Universal}
When it is possible to perform a universal set of gates, it suffices
to prepare the $|0\>$ state and to measure along the computation basis
$\{|0\>,|1\>\}$ in conditions (1) and (3).


We describe some universal sets of gates that are relevant to the
current discussion.  The first one is the set of all $2$-qubit
gates.\cite{DiVincenzo95b} The second set consists of all $1$-qubit
gates and only one $2$-qubit gate, the controlled-{\sc not} ({\sc
cnot}).\cite{Barenco95a} These two continuous sets of gates generate
any unitary operation {\em exactly}.  There are also finite, discrete
universal sets that generate any operation to arbitrary accuracy.  For
example, the {\sc cnot}, the phase gate, the Hadamard gate, and the
$\pi \over 8$ gate form a universal set.\cite{Boykin99} Also, almost
any single $2$-qubit gate is
universal.\cite{Lloyd95a,Deutsch95a,Barenco95b}

Other computation models have been built upon the standard model, so
as to achieve fault tolerance or to adapt to promising physical
systems.  In these models, only some unitary gates can be easily
performed and they do not form a universal set.
In the context of fault tolerant quantum computation, Shor pioneered a
recipe that {\em indirectly} performs the Toffoli gate using an ancilla,
measurements, and some other gates.\cite{Shor96} 
The method was generalized\cite{Gottesman99,Zhou00} by
understanding the connection to teleportation.\cite{Bennett93}
The generalized method was applied to quantum computation schemes 
based on linear optics\cite{Knill01} and exchange interactions\cite{Lidar02}.  
%

More recently, two different models of quantum computation based only
on measurements are proposed.
Raussendorf and Briegel\cite{Raussendorf01} have proposed a ``$1$-way
quantum computer'' which starts with a {\em cluster
state}\cite{Briegel00} of certain size and uses only $1$-qubit
measurements.
The cluster state can be replaced by a circuit-dependent initial state
that takes $4$-qubit measurements to prepare.
Independently, Nielsen\cite{Nielsen01p} extended the indirect methods
of performing unitary gates to obtain a quantum computation model 
using only measurements on up to $4$ qubits.  
%

The central result in Ref.~\refcite{Nielsen01p} is a method to perform
the universal set of all two-qubit gates using $4$-qubit measurements
only.
Focusing on the smaller set of $1$-qubit gates and the {\sc cnot},
Fenner and Zhang\cite{Fenner01} and independently Leung and
Nielsen\cite{Leung01mit} showed that $3$-qubit measurements are
universal.
Subsequently, Leung\cite{Leung01} showed that $2$-qubit measurements
are sufficient.  This is the minimal number of qubits to be measured
jointly in order to achieve universality because measurements are the
only means of interaction.
This result parallels the universality of $2$-qubit
gates\cite{DiVincenzo95b} in the standard model.


In this paper, we simplify and extend the results in
Refs.~\refcite{Nielsen01p} and \refcite{Leung01}.
We systematically consider indirect implementation of unitary gates,
including the method proposed in Ref.~\refcite{Gottesman99} and a new
method that partially collapses the hierarchy of unitary gates
proposed in Ref.~\refcite{Gottesman99}.
We proceed to first rederive the universality of $4$-qubit
measurements\cite{Nielsen01p} based on a much simpler scheme.
Second, we rederive the result in Ref.~\refcite{Leung01} that
$2$-qubit measurements are universal.  The current construction
differs from Ref.~\refcite{Leung01} in using deterministic resources,
similar to the $1$-way quantum computer\cite{Raussendorf01}. 
%
%
Third, we prove that a single $4$-qubit measurement can be universal,
and we prove an analogue of the result that almost any 
$2$-qubit gate is universal.\cite{Barenco95b,Lloyd95a,Deutsch95a}

\section{Indirect Implementations of Quantum Gates}

We first review the Pauli and Clifford groups (see
Refs.~\refcite{Gottesman97,Nielsen00,Leung00} for example).
Let $\sigma_{1,2,3}$ or $X$, $Y$, $Z$ be the Pauli operators and
$\sigma_0$ be the $2 \times 2$ identity matrix, 
\be
\sigma_0 = \left( \begin{array}{cc} 1 & 0 \\ 0 & 1 \end{array} \right) ,\, 
\sigma_1 = \left( \begin{array}{cc} 0 & 1 \\ 1 & 0 \end{array} \right) ,\,
\sigma_2 = \left( \begin{array}{cc} 0 &-i \\ i & 0 \end{array} \right) ,\,
\sigma_3 = \left( \begin{array}{cc} 1 & 0 \\ 0 &-1 \end{array} \right) .
\ee
The Pauli group is generated by $\sigma_j$ acting on each qubit.  The
Clifford group consists of unitary operators that conjugate Pauli
operators to Pauli operators.  The Clifford group is generated by the
{\sc cnot}, the phase gate $\mbox{\sc p} = e^{-i {\pi \over 4} Z}$,
and the Hadamard gate $\mbox{\sc h} = \os2 (X+Z) $.

A crucial element in performing unitary gates indirectly is
teleportation\cite{Bennett93} that transmits a qubit $|\psi\> = a |0\>
+ b |1\>$ using the following circuit:
\bea
\mbox{\psfig{file=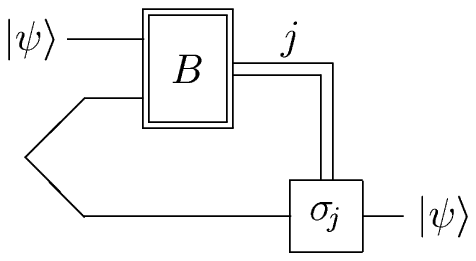,width=1.55in}}
\label{eq:teleport} 
\eea
In the circuits throughout this paper, time goes from left to right,
single lines denote qubits, double lines denote classical data,
single- and double-lined boxes respectively denote unitary gates and
measurements.  
A gate that is connected to a measurement box by a double line is 
performed conditioned on the measurement outcome.  
Two qubits connected in the left are initially in a maximally entangled state 
$\B{0} = \os2 (|00\> + |11\>)$. 
The Bell measurement, labeled $B$, is along the Bell basis: 
\bea
	\B{0} & = &\os2 (|00\> + |11\>) \,,~~
	\B{3} = \os2 (|00\> - |11\>) \,,
\non
\\
	\B{1} & = & \os2 (|01\> + |10\>) \,,~~
	\B{2} = \os2 (|01\> - |10\>) \,.
\label{eq:bell}
\eea
Equation (\ref{eq:teleport}) can be verified by rewriting the initial
state $|\psi\> \B{0}$ as 
${1 \over 2} \sum_j \B{j} \! \ot \! (\sigma_j |\psi\>)$.
%
%
The Bell measurement on the first two qubits collapses the initial
state to one of the four terms, and conditioned on the outcome $j$, 
$\s_j$ is applied on the last qubit to recover $|\psi\>$.

Teleportation was initially proposed as a communication protocol, but
it turns out to be useful in indirect gate constructions.  
In particular, there are two trivial methods to apply a gate $U$ to a
state $|\psi\>$.  
One can teleport $|\psi\>$ and then apply $U$: 
\bea
\mbox{\psfig{file=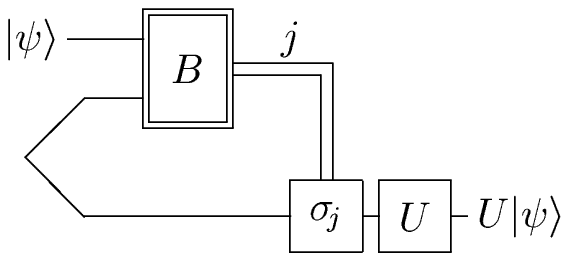,width=1.7in}}
\label{eq:homersimpson}
\eea
which implies the validity of the following circuit:
\bea
\mbox{\psfig{file=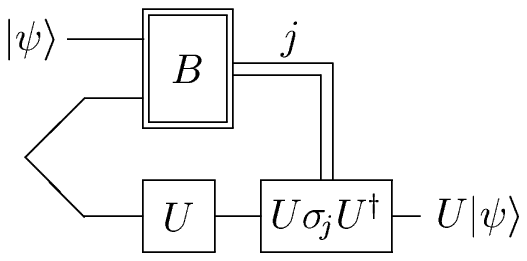,width=1.7in}}
\label{eq:heart}
\eea
Throughout the paper, $\dagger$, $T$, and $*$ denote the adjoint, the
transpose, and the complex conjugate of an operator respectively.
Note that the gate $U \s_j U^\dagger$ is the same as the gates 
$U^\dagger$, $\s_j$, and $U$ applied in order.
Equation (\ref{eq:heart}) is a recipe for an indirect implementation
of $U$ by preparing an ``ancilla'' $\os2 (I \ot U)(|00\>+|11\>)$ and
applying a Bell measurement and a ``correction'' $U \s_{j} U^\dag$.
It is indirect in that, one must provide the ancilla 
and the correction 
with resources allowed in the context, without applying $U$ (see
Refs.~\refcite{Shor96,Boykin99,Gottesman99,Zhou00,Knill01,Nielsen01p}
in the contexts of fault-tolerance and alternative computation
models).

The second trivial method to apply $U$ to $|\psi\>$ is to apply $U$
and teleport $U |\psi\>$:\cite{newtrick}
\bea
\mbox{\psfig{file=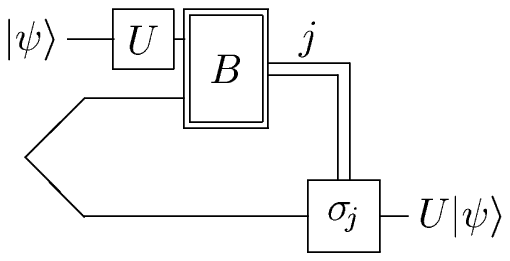,width=1.6in}}
\label{eq:gottesman}
\eea
Applying $U$ on the first register followed by the Bell measurement is
equivalent to applying a measurement $B_{U^\dagger_1}$ along the basis
$\{ (U^\dagger \ot I) \B{j} \}_j$: 
%
%
\bea
\mbox{\psfig{file=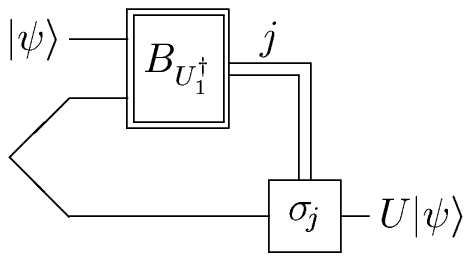,width=1.45in}}
\label{eq:bu}
\eea

Equations (\ref{eq:heart}) and (\ref{eq:bu}) also generalize trivially
to any $n$-qubit gate.  We focus on the $2$-qubit generalizations:
\bea
\setlength{\unitlength}{0.6mm}
\centering
\begin{picture}(68,45)
%
\put(-10,35){\makebox(10,10){$\<2a\>$}}
\put(0,32.5){\makebox(12,10){$|\psi\rangle$}}

\put(12,40){\line(1,0){8}}
\put(12,35){\line(1,0){8}}


\put(15,30){\line(1,0){5}}
\put(15,25){\line(1,0){5}}

\put(5,15){\line(1,1){10}}
\put(5,15){\line(1,-1){10}}
\put(5,20){\line(1,1){10}}
\put(5,20){\line(1,-1){10}}
{\small 
\put(37,20){\makebox(10,10){$j$}}  
\put(20,22.5){\framebox(12,20){$B^{\!\otimes\! 2}$}}
\put(20.7,23.2){\framebox(10.6,18.6){}} }
\put(32,33){\line(1,0){7}}
\put(32,32){\line(1,0){6}}
\put(39,33){\line(0,-1){20}}
\put(38,32){\line(0,-1){19}}
{\small  \put(28,03){\framebox(20,10){$U \! P_j U^\dagger$}}  }
\put(15,05){\line(1,0){3}}
\put(15,10){\line(1,0){3}}
{\small  \put(18,03){\framebox(7,10){$U$}}  }
\put(25,05){\line(1,0){3}}
\put(25,10){\line(1,0){3}}
\put(48,05){\line(1,0){5}}
\put(48,10){\line(1,0){5}}
\put(54,02){\makebox(13,10){$U|\psi\rangle$}}
\end{picture}
~~~~~~~~~~
\begin{picture}(70,45)
%
\put(-10,35){\makebox(10,10){$\<2b\>$}}
\put(0,32.5){\makebox(12,10){$|\psi\rangle$}}

\put(12,40){\line(1,0){13}}
\put(12,35){\line(1,0){13}}

\put(15,30){\line(1,0){10}}
\put(15,25){\line(1,0){10}}

\put(5,15){\line(1,1){10}}
\put(5,15){\line(1,-1){10}}
\put(5,20){\line(1,1){10}}
\put(5,20){\line(1,-1){10}}
{\small 
\put(42,20){\makebox(10,10){$j$}}  
\put(25,22.5){\framebox(12,20){$B_{\!U\!^\dagger_{1\!2}}$}}
\put(25.7,23.2){\framebox(10.6,18.6){}} }
\put(37,33){\line(1,0){7}}
\put(37,32){\line(1,0){6}}
\put(44,33){\line(0,-1){20}}
\put(43,32){\line(0,-1){19}}
{\small  \put(38,03){\framebox(10,10){$P_j$}}  }
\put(15,05){\line(1,0){8}}
\put(15,10){\line(1,0){8}}
\put(23,05){\line(1,0){10}}
\put(23,10){\line(1,0){10}}
\put(33,05){\line(1,0){5}}
\put(33,10){\line(1,0){5}}
\put(48,05){\line(1,0){5}}
\put(48,10){\line(1,0){5}}
\put(56,02){\makebox(13,10){$U|\psi\rangle$}}
\end{picture}
%
%
\label{eq:heart2}
\eea
In \eq{heart2} $\<2a\>$, $B^{\ot 2}$ stands for $B_{13} \ot B_{24}$, a
Bell measurement on qubits $1,3$ and one on qubits $2,4$.
In \eq{heart2} $\<2b\>$, we have combined $U$ and $B^{\ot 2}$ in 
a single measurement $B_{U^\dagger_{12}}$ along the basis $\{
(U_{12}^\dagger \ot I_{34}) (\B{{j_1}}_{13} \ot \B{{j_2}}_{24}) \}_{j_1,j_2}$.
We denote a $2$-qubit Pauli operator as $P_j = \s_{j_1} \ot \s_{j_2}$ 
where $j \equiv (j_1,j_2)$.
The following variants of Eqs.~(\ref{eq:heart}) and (\ref{eq:bu}) will
also be useful:
\bea 
\setlength{\unitlength}{0.6mm}
\centering
\begin{picture}(170,40)
\put(-4,30){\makebox(10,10){$\<1a\>$}} 
\put(6,0){\psfig{file=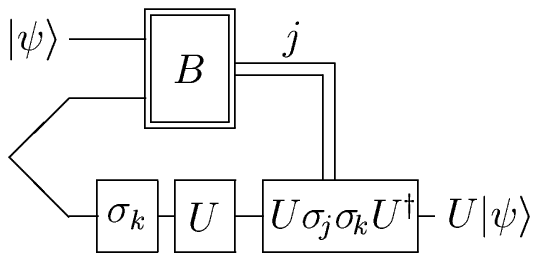,width=1.75in}}
\put(85,30){\makebox(10,10){$\<1b\>$}} 
\put(95,0){\psfig{file=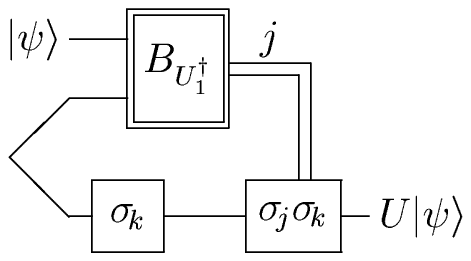,width=1.50in}}
\end{picture}
\label{eq:heartvar}
\eea

%
%
%
%

\section{Quantum Computation by measurements only} 

Given the ability to perform projective measurements, state
initialization and final readout are trivial, and according to the
standard model, it remains to perform a universal set of gates.  We
will explain in detail how to use measurements to provide the various
resources needed in the schemes $\<1a,b\>$ and $\<2a,b\>$ in
\eqs{heart2}{heartvar} to perform nontrivial unitary operations that
form a universal set.
The identity and {\sc swap} operations are implicitly provided by
quantum storage and by the ability to choose which qubit to measure.

We first demonstrate how to perform a Pauli operation $\sigma_l$ using
Bell measurements only.  This illustrates some of the basic ideas, and
the Pauli operations will also be used repeatedly in subsequent
discussions.  Consider method $\<1b\>$ in \eq{heartvar}.  The ancilla
can be taken to be $(I \otimes \sigma_k) \B{0} = \B{k}$ for any $k$,
and can be obtained as a post-measurement state of a single Bell
measurement on any $2$-qubit system.  When $U=\sigma_j$,
$B_{U_1^\dagger}$ along the basis $\{(U^\dagger \ot I)\B{j}\}$ is just
the Bell measurement (outcomes redefined).  With probability $1/4$, $j
= k$ and no correction gate is needed, in which case the desired
$\sigma_l$ is performed with $2$ Bell measurements.
Otherwise, the Pauli correction can be performed {\em recursively},
again completed with probability $1/4$ when no further correction is
needed.
One repeats the recursion until correction is unnecessary, which on
average occurs after $4$ trials and $8$ Bell measurements.
This performs a deterministic Pauli operation with variable resources. 

\subsection{Universality of $4$-qubit measurements}

We now give a simple derivation that $4$-qubit measurements are
universal\cite{Nielsen01p} by using them to perform any $2$-qubit gate
$U$ via method $\<2b\>$ in \eq{heart2}.  We use Bell measurements to
perform the Pauli correction and to provide the ancilla, which can be
any two Bell states by redefining the correction operation similar to
method $\<1b\>$ in \eq{heartvar}.  Finally, $B_{U_{12}^\dagger}$ is a
$4$-qubit measurement directly available.

\subsection{Universality of $2$-qubit measurements}

We now demonstrate the universality of $2$-qubit
measurements.\cite{Leung01}
To do this, we consider the simpler universal set of all $1$-qubit
gates and the {\sc cnot}.  
The $1$-qubit gates can be performed using method $\<1b\>$ in
\eq{heartvar}, which requires $2$-qubit measurements only.
We use method $\<2a\>$ to perform {\sc cnot} to avoid applying the
unavailable $4$-qubit measurement $B_{U^\dagger_{12}}$ to the input
state.  This comes at the cost of two extra complications.  First, the
correction gate becomes $\mbox{\sc cnot} \; (\sigma_{j_1} \ot
\sigma_{j_2}) \; \mbox{\sc cnot}$, but this is just a tensor product
of Pauli operators since {\sc cnot} is in the Clifford group.  The
second complication and the last obstacle is the need to obtain the
special ancilla,
\bea
	|a_{\rm cn}\> 
	& = & {1 \over 2} (I \ot I \ot \mbox{\sc cnot})
	(|0000\> + |0101\> + |1010\> + |1111\>)
\non
\\
	& = & {1 \over 2} 
	(|0000\> + |0101\> + |1011\> + |1110\>) \,. 
\label{eq:acn}
\eea
We prove the universality of $2$-qubit projective measurements by
showing how they can be used to prepare $|a_{\rm cn}\>$.
For simplicity, we focus on measurement outcomes that correspond to 
the postmeasurement state 
$|a_{\rm cn}\>$.  We will see later other outcomes
correspond to equally good ancillas.
We first present the method in the state representation: \\[1.2ex]
1. Create ${1 \over 2} (|0\>+|1\>) \ot |0\> \ot (|00\>+|11\>)$
with $1$- and $2$-qubit measurements.  \\[1.2ex]
2. Apply to qubits 2,3 the measurement with 2 projectors: 
%
\bea 
	P_+ & = & \B{0} \<\Phi_0| + |\Phi_1\>\<\Phi_1|
	= {1 \over 2} (|00\> \! + \! |11\>) (\<00|\!+\!\<11|) 
	+ {1 \over 2} (|01\>\!+\!|10\>) (\<01|\!+\!\<10|) \,,
\non
\\
	P_- & = & |\Phi_2\>\<\Phi_2| + |\Phi_3\>\<\Phi_3|
	= {1 \over 2} (|00\>\!-\!|11\>) (\<00|\!-\!\<11|) 
	+ {1 \over 2} (|01\>\!-\!|10\>) (\<01|\!-\!\<10|) \,.
\non
\eea
When the outcome corresponds to $P_+$, the state becomes ${1 \over 2
\sqrt{2}} (|0\>+|1\>) \ot (|000\> + |011\> + |101\> + |110\>)$. \\[1.2ex]
3. Measure the parity of qubits 1,3.  If the outcome is even, the
state becomes
${1 \over 2}(|0000\> + |1011\> + |0101\> + |1110\>)$, 
which is $|a_{\rm cn}\>$. 
%

We can also explain the above scheme in the stabilizer
language.\cite{Gottesman97,Gottesman98} 
The stabilizer of an $n$-qubit state $|\psi\>$ is an abelian group
with $n$ generators $O_i$ such that $O_i |\psi\> = |\psi\>$.
These generators specify the state up to a phase.  
If $O |\psi\> = |\psi\>$, $U O U^\dagger (U |\psi\>) = U|\psi\>$,
therefore, the state evolves as $|\psi\> \rightarrow U|\psi\>$ when
each generator evolves as $O \rightarrow U O U^\dagger$.
Furthermore, suppose $M$ is a traceless operator with eigenvalues $\pm
1$, and it commutes or anticommutes with each generator.
If the outcomes $\pm 1$ are obtained when measuring $M$, the
generators that anticommute with $M$ evolve as $\{N_1, N_2, N_3,
\cdots\} \rightarrow \{\pm M, N_1 N_2, N_1 N_3, \cdots\}$.

The stabilizer of $\B{0}_{1,3} \ot \B{0}_{2,4}$ is generated by
$XIXI$, $ZIZI$, $IXIX$, $IZIZ$, where $XIXI = \sigma_x \ot I \ot
\sigma_x \ot I$ and so on.  Since 
%
%
%
%
\bea
	& \mbox{\sc cnot} \, (X I) \, \mbox{\sc cnot} = X \!X \,, &  
	\mbox{\sc cnot} \, (I X) \, \mbox{\sc cnot} = IX \,, 
\\
	& \mbox{\sc cnot} \, (Z I) \, \mbox{\sc cnot} = ZI \,, & 
	\mbox{\sc cnot} \, (I Z) \, \mbox{\sc cnot} = ZZ \,, 
\eea
the stabilizer of $|a_{\rm cn}\>$ is generated by: 
\bea
	XI XX ,\; ZI ZI ,\; IX IX ,\; IZ ZZ \,.
\label{eq:ancgen}
\eea
One can prepare a state by measuring the generators of its stabilizer.
However, any generator set for $|a_{\rm cn}\>$ contains elements of
weight $3$ (the weight is the number of nontrivial tensor components).
Our strategy is to start with an initial state with generators of
weights $1$ and $2$ (step 1) and apply $2$-qubit measurements $IXXI$
and then $ZIZI$ to {\em induce} multiplications between generators
that anticommute with the measured operator, thereby increasing the
weights of the generators.
Assuming $+1$ outcomes, the evolution is given by:
\bea
\non
	\begin{array}{cc}
	XI & II
\\	IZ & II 
\\	II & XX 
\\	II & ZZ 
	\end{array}
~~~
	\stackrel{\begin{array}{c}{\rm measure}
          \\{IXXI}\end{array}}{\longrightarrow} 
~~~
	\begin{array}{cc}
	XI & II
\\	IX & XI 
\\	II & XX
\\	IZ & ZZ
	\end{array}
~~~
	\stackrel{\begin{array}{c}{\rm measure}
          \\{ZIZI}\end{array}}{\longrightarrow} 
~~~
	\begin{array}{cc}
	ZI & ZI
\\	XX & XI
\\	XI & XX
\\	IZ & ZZ
	\end{array} 	
\stackrel{\begin{array}{c}{}
          \\{.}\end{array}}{} 
\eea
The final set of generators is equivalent to that in \eq{ancgen}
because multiplying one generator to another does not affect the stabilizer.

We have focused on measurement outcomes that result in $|a_{\rm
cn}\>$ in the above discussion.
Other outcomes result in states of the form $(\s_{k} \ot
\s_{l} \ot \mbox{\sc cnot}) \B{0}_{1,3} \ot \B{0}_{2,4} = \pm \lpm I
\ot I \ot (\mbox{\sc cnot} \; \s_{k} \ot \s_{l}) \rpm \B{0}_{1,3} \ot
\B{0}_{2,4}$ (which can be used as the ancilla by adapting the
correction procedure as in \eq{heartvar}).
This is immediate in the stabilizer representation -- post-measurement
states of other outcomes differ by extra $-$ signs in some of the
generators.
These signs can be induced by applying Pauli operators to the first
$2$ qubits of $|a_{\rm cn}\>$.  Thus other output states are precisely
$(\s_{k} \ot \s_{l} \ot II )|a_{\rm cn}\>$.
This can also be verified directly in the state representation.

We turn our attention to discrete universal sets of (incomplete)
$2$-qubit measurements that correspond to discrete universal sets of
gates.
It is known that the Clifford group generated by $\{${\sc cnot}, {\sc
h}, {\sc p}$\}$ together with any other gate are
universal~\cite{Gottesmanpriv}.  Thus $\{${\sc cnot}, {\sc h}, {\sc
p}, {\sc u}$\}$ is universal for any 1-qubit gate {\sc u} outside the
Clifford group.
We can enumerate all the required measurements.
First, all correction gates are Pauli operators, requiring only Bell
measurements (i.e.~measuring $XX$ and $ZZ$).
We need to perform $B_{\mbox{\sc h}^\dagger_1}$, $B_{\mbox{\sc
u}^\dagger_1}$, and $B_{\mbox{\sc p}^\dagger_1}$ for the $1$-qubit
gates. 
In general, $B_{U^\dagger_1}$ is a measurement of the operators 
$(U^\dagger X U) \ot X$ and $(U^\dagger Z U) \ot Z$.
Thus we need to measure $XZ$, $XY$, and $(\mbox{\sc u}^\dagger X
\mbox{\sc u}) \ot X$ and $(\mbox{\sc u}^\dagger Z \mbox{\sc u}) \ot Z$
to perform {\sc h}, {\sc p}, and {\sc u} respectively.
Finally, ancilla preparation requires other measurements.  The Bell
state ancilla for any $1$-qubit gate requires no new measurement.  The
preparation of the state $|a_{\rm cn}\>$ for {\sc cnot} requires the
states $|0\>$ and $\os2 (|0\>+|1\>)$.  We can measure $Z$ to prepare
$|0\>$, and apply {\sc h} to $|0\>$ to obtain $\os2 (|0\>+|1\>)$. 
%
%
%
%
Altogether, 
\bea
	S_0 = \!\! 
	\{Z, XX, ZZ, XZ, XY, (\mbox{\sc u}^\dagger X \mbox{\sc u}) \ot X, 
	(\mbox{\sc u}^\dagger Z \mbox{\sc u}) \ot Z \}
\non
\eea
is universal. 
Special choices of {\sc u} can further simplify the universal set.  
For instance, 
\bea
S_1 & = \! & \{Z, XX, ZZ, XZ, XY, 
(\cos \theta \, Z + \sin \theta \, Y) \ot Z \}
\non
\\ 
S_2 & = \! & \{Z, XX, ZZ, XZ, XY, 
(\cos \theta \, X + \sin \theta \, Y) \ot X \}
\non
\\
S_3 & = \! & \{Z, XX, ZZ, XZ, \os2 (X\!+\!Y) \ot X \}
\non
\eea 
are universal sets of measurements corresponding to {\sc u}
$= e^{i {\theta \over 2} X}$, $e^{-i {\theta \over 2} Z}$, and
$e^{-i {\pi \over 8} Z}$ respectively ($\theta \neq m \pi/2$ for
$m$ an integer).
$S_3$ corresponds to the universal set of gates $\{e^{-i {\pi \over 8}
Z},$ {\sc h}, {\sc cnot}$\}$,
%
%

The above scheme implements a desired gate precisely at each stage of
a computation by performing the Pauli correction with variable
resources.  However, this is unnecessary.  Suppose the sequence of
gates $\{V_1, V_2, \cdots\}$ are to be applied to the $i$-th qubit,
where $V_j$ is a 1-qubit gate or a {\sc cnot} (involving another
qubit).  Instead of applying the correction $\sigma_{i_1}$ for $V_1$,
we can {\em absorb} $\sigma_{i_1}$ into the next gate $V_2$, i.e., to
perform $V_2 \; \sigma_{i_1}$ instead.
If $V_2$ is in the Clifford group, then, $V_2 \, \sigma_{i_1} =
\sigma_{i_1'} \, V_2$ for some $i_1'$ and we simply perform $V_2$.  If
the correction for $V_2$ is $\sigma_{i_2}$, the combined correction is
$\sigma_{i_1'} \sigma_{i_2}$.  Thus the correction for $V_1$ can be
omitted by redefining the correction for $V_2$.  The combined
$\sigma_{i_1'} \sigma_{i_2}$ can now be absorbed in $V_3$ similarly.
If $V_2$ is not in the Clifford group, then $V_2 = \mbox{\sc u}$ is a
$1$-qubit gate, and we perform $\mbox{\sc u} \sigma_{i_1}$ by
replacing $B_{{\mbox{\sc u}}_1^\dagger}$ in method $\<1b\>$
(\eq{heartvar}) by $B_{{(\sigma_{i_1} \mbox{\sc u})}_1^\dagger}$.
Now, our universal sets of measurements $S_{1,2,3}$ each require
an extra element similar to the last element listed for each set.

\subsection{Universality of a single $4$-qubit measurement}

We turn to a different task that parallels the search of a single
$2$-qubit gate that is
universal.\cite{Barenco95b,Lloyd95a,Deutsch95a}.
We will show that the $4$-qubit measurement $B_{U^\dagger_{12}}$ alone
is universal for appropriately chosen $U$.
We denote the Pauli group over $2$ qubits by ${\cal P}_2$, with 
elements $P_j = \s_{j_1} \ot \s_{j_2}$, $j \equiv (j_1,j_2)$.
We first find out what gates can be performed by using the measurement
$B_{U^\dagger_{12}}$.  According to \eq{heart2}, the following 
circuits are valid: 
\begin{equation}
\setlength{\unitlength}{0.6mm}
\centering
\begin{picture}(70,46)
%
\put(-12,35){\makebox(10,10){(A)}}
\put(0,32.5){\makebox(12,10){$|\psi\rangle$}}

\put(12,40){\line(1,0){13}}
\put(12,35){\line(1,0){13}}

\put(15,30){\line(1,0){10}}
\put(15,25){\line(1,0){10}}

\put(5,15){\line(1,1){10}}
\put(5,15){\line(1,-1){10}}
\put(5,20){\line(1,1){10}}
\put(5,20){\line(1,-1){10}}
{\small 
\put(42,20){\makebox(10,10){$j$}}  
\put(25,22.5){\framebox(12,20){$B_{\!U\!^\dagger_{1\!2}}$}}
\put(25.7,23.2){\framebox(10.6,18.6){}} }
\put(37,33){\line(1,0){7}}
\put(37,32){\line(1,0){6}}
\put(44,33){\line(0,-1){15}}
\put(43,32){\line(0,-1){14}}
\put(15,05){\line(1,0){8}}
\put(15,10){\line(1,0){8}}
\put(23,05){\line(1,0){10}}
\put(23,10){\line(1,0){10}}
\put(33,05){\line(1,0){5}}
\put(33,10){\line(1,0){5}}
\put(40,02){\makebox(20,10){$P_j U|\psi\rangle$}}
\end{picture}
%
%
~~~~~~
\begin{picture}(70,46)
%
\put(-12,35){\makebox(10,10){(B)}}
\put(0,32.5){\makebox(12,10){$|\psi\rangle$}}

\put(12,40){\line(1,0){13}}
\put(12,35){\line(1,0){13}}

\put(15,30){\line(1,0){10}}
\put(15,25){\line(1,0){10}}

\put(5,15){\line(1,1){10}}
\put(5,15){\line(1,-1){10}}
\put(5,20){\line(1,1){10}}
\put(5,20){\line(1,-1){10}}
{\small 
\put(42,20){\makebox(10,10){$j$}}  
\put(25,22.5){\framebox(12,20){$B_{\!U\!^\dagger_{3\!4}}$}}
\put(25.7,23.2){\framebox(10.6,18.6){}} }
\put(37,33){\line(1,0){7}}
\put(37,32){\line(1,0){6}}
\put(44,33){\line(0,-1){15}}
\put(43,32){\line(0,-1){14}}
\put(15,05){\line(1,0){8}}
\put(15,10){\line(1,0){8}}
\put(23,05){\line(1,0){10}}
\put(23,10){\line(1,0){10}}
\put(33,05){\line(1,0){5}}
\put(33,10){\line(1,0){5}}
\put(40,02){\makebox(20,10){$U^T P_j |\psi\rangle$}}
\end{picture}
\label{eq:newnofix}
\end{equation}

Circuit (A) is derived from method $\<2b\>$ in \eq{heart2} without
correcting for $P_j$.  In circuit (B) $B_{\!U\!^\dagger_{3\!4}}$
denotes the same measurement as $B_{\!U\!^\dagger_{1\!2}}$ with qubits
$(1,2)$ and $(3,4)$ interchanged.  This measurement has the same
effect as $U$ acting on qubits 3,4 followed by Bell measurements
$B_{13} \ot B_{24}$.  The circuit follows from teleportation and the
fact that $U$ acting on half of a maximally entangled state is the
same as $U^T$ acting on the other half.

Bell states used in \eq{newnofix} are not directly available
in the current task.  Instead, we use the following ancillas 
obtainable from $B_{\!U\!^\dagger_{1\!2}}$: 

\begin{equation}
\setlength{\unitlength}{0.6mm}
\centering
~~
\begin{picture}(120,40)
\put(-5,37){\makebox{(C)}}
\put(00,30){\line(1,0){5}}
\put(00,25){\line(1,0){5}}
\put(00,20){\line(1,0){5}}
\put(00,15){\line(1,0){5}}

\put(17,30){\line(1,0){5}}
\put(17,25){\line(1,0){5}}
\put(17,20){\line(1,0){5}}
\put(17,15){\line(1,0){5}}

{\small 
\put(05,12.5){\framebox(12,20){$B_{\!U\!^\dagger_{1\!2}}$}}
\put(05.7,13.2){\framebox(10.6,18.6){}} }

\put(11,05){\makebox(6,10){$k$}}  

\put(10.5,12.5){\line(0,-1){5}}
\put(11.5,12.5){\line(0,-1){5}}

\put(26,18){\makebox(6,10){$=$}}  

\put(35,20){\line(1,1){10}}
\put(35,20){\line(1,-1){10}}
\put(35,25){\line(1,1){10}}
\put(35,25){\line(1,-1){10}}

\put(45,15){\line(1,0){25}}
\put(45,10){\line(1,0){25}}

\put(45,35){\line(1,0){5}}
\put(45,30){\line(1,0){5}}
\put(50,27.5){\framebox(8,10){$P_k$}}  
\put(58,35){\line(1,0){2}}
\put(58,30){\line(1,0){2}}
\put(60,27.5){\framebox(8,10){$U^\dagger$}}  
\put(68,35){\line(1,0){2}}
\put(68,30){\line(1,0){2}}

\put(73,18){\makebox(6,10){$=$}}  

\put(82,20){\line(1,1){10}}
\put(82,20){\line(1,-1){10}}
\put(82,25){\line(1,1){10}}
\put(82,25){\line(1,-1){10}}

\put(092,35){\line(1,0){25}}
\put(092,30){\line(1,0){25}}

\put(092,15){\line(1,0){5}}
\put(092,10){\line(1,0){5}}

\put(097,07.5){\framebox(8,10){$U^*$}}  
\put(105,15){\line(1,0){2}}
\put(105,10){\line(1,0){2}}
\put(107,07.5){\framebox(8,10){$P_k$}}  
\put(115,15){\line(1,0){2}}
\put(115,10){\line(1,0){2}}
\end{picture}
%
%
~~~~~~~~~~~~~~~
\begin{picture}(73,43)
\put(-5,37){\makebox{(D)}}
\put(00,30){\line(1,0){5}}
\put(00,25){\line(1,0){5}}
\put(00,20){\line(1,0){5}}
\put(00,15){\line(1,0){5}}

\put(17,30){\line(1,0){5}}
\put(17,25){\line(1,0){5}}
\put(17,20){\line(1,0){5}}
\put(17,15){\line(1,0){5}}

{\small 
\put(05,12.5){\framebox(12,20){$B_{\!U\!^\dagger_{3\!4}}$}}
\put(05.7,13.2){\framebox(10.6,18.6){}} }

\put(11,05){\makebox(6,10){$k$}}  

\put(10.5,12.5){\line(0,-1){5}}
\put(11.5,12.5){\line(0,-1){5}}

\put(26,18){\makebox(6,10){$=$}}  

\put(35,20){\line(1,1){10}}
\put(35,20){\line(1,-1){10}}
\put(35,25){\line(1,1){10}}
\put(35,25){\line(1,-1){10}}

\put(45,35){\line(1,0){25}}
\put(45,30){\line(1,0){25}}

\put(45,15){\line(1,0){5}}
\put(45,10){\line(1,0){5}}

\put(50,07.5){\framebox(8,10){$P_k$}}  
\put(58,15){\line(1,0){2}}
\put(58,10){\line(1,0){2}}
\put(60,07.5){\framebox(8,10){$U^\dagger$}}  
\put(68,15){\line(1,0){2}}
\put(68,10){\line(1,0){2}}
\end{picture}
\label{eq:anc2}
\end{equation}

In particular, we can perform some gates using the following two primitives :
\begin{enumerate}
\item If ancilla (D) is used in place of the Bell states in circuit
(A), the output state will be $U^\dagger P_k P_j U|\psi\>$.  Thus
one can apply $U^\dagger P_k P_j U$ to any $2$-qubit state.
\item Likewise, replacing the Bell states in circuit (B) by ancilla
(C), the final state is $P_k U^* U^T P_j |\psi\> = P_k P_j |\psi\>$,
thus $P_k P_j$ is applied to the input state.
\end{enumerate}
Using primitive (2), a random element from the set $\{P_k\}_k$ can be
performed.  The randomness can be removed.  Repeating primitive (2)
$l$ times results in the gate $G_l = P_{k_l} \cdots P_{k_2} P_{k_1}
\in {\cal P}_2$.  $\{G_l\}$ is a random walk on ${\cal P}_2$ that hits
any desired $P \in {\cal P}_2$ with an average of $16$ iterations.
Likewise, repeated use of primitive (1) results in a similar random
walk $U^\dagger P_{k_l} \cdots P_{k_2} P_{k_1} U$ allowing $U^\dagger
P U$ to be performed for any desired $P$.

In particular, $Q U^\dagger P U$ can be performed for any $Q,P \in
{\cal P}$ of our choice.  We claim that $Q U^\dagger P U$ is
universal when $P = Q = I \ot Z$,  
\bea 
	U = \left[ \begin{array}{c|c}
	I_2 & 0_2  
\\ \hline	
	0_2 & R
	\end{array}
	\right]
\eea
where $I_2, 0_2$ are the $2 \times 2$ identity and zero matrices 
respectively, and 
\bea
	R = \left[ \begin{array}{lr}
	\cos \theta & \!\!\!\!\! -i e^{-i \phi} \sin \theta  
\\	-i e^{i \phi} \sin \theta & \!\!\!\!\! \cos \theta
	\end{array}
	\right]
\eea
is a rotation of $2 \theta$ about the axis $\cos \phi \;
X + \sin \phi \; Y$.  
Then, 
\bea
	& & Q U^\dagger P U  
	= 
	\left[ \begin{array}{c|c} Z & 0_2  \\ \hline 0_2 & Z \end{array}
	\right] \times 
	\left[ \begin{array}{c|c} I_2 & 0_2  \\ \hline 0_2 & R^\dag \end{array}
	\right] \times 
	\left[ \begin{array}{c|c} Z & 0_2  \\ \hline 0_2 & Z \end{array}
	\right] \times 
	\left[ \begin{array}{c|c} I_2 & 0_2  \\ \hline 0_2 & R \end{array}
	\right] 
	= 
	\left[ \begin{array}{c|c} I_2 & 0_2  \\ 
		\hline 0_2 & (Z R^\dag Z) R \end{array} \right] .  
\non
\eea
Note that $Z R^\dag Z = R$ because $Z$ anticommutes with both
$X$ and $Y$ and reverses $R^\dag$.
Thus, $Q U^\dagger P U = U^2$, which is universal if $\theta$
and $\phi$ are both irrational multiples of $\pi$, and their ratio is
also irrational.\cite{Barenco95b} 

The measurement $B_{U^\dagger_{12}}$ is actually universal for almost
all $2$-qubit gates $U$.  This is because the spectra of $H$ where
$e^{-iH} = Q U^\dagger P U$ form a set of positive measure in $\RR^4$
when $P,Q,U$ are varied, while the set of all nonuniversal $2$-qubit
gates is of zero measure.\cite{Lloyd95a,Deutsch95a} On the other hand,
denote the Clifford group by $C_2$ and the set that conjugates the
Pauli group into the Clifford group by $C_3$.\cite{Gottesman99} If $U
= U_1 U_2^\dagger$ for $U_1,U_2 \in C_3$.  Our scheme only generates
$P_1 U^\dagger P_2 U P_3 \cdots$ which is always in $P U_2 C_2
U_2^\dagger Q$.

\section{Conclusions}

We described a variety of methods for performing gates by measurements
only.
We showed that $2$-qubit measurements are necessary and sufficient for
universal quantum computation.  This is optimal in the number of
qubits to be measured jointly.
We proved the universality of almost all single maximally entangling
$4$-qubit measurements.  This is minimal in the number of measurements
available.
Method $\<1b\>$ (and its $n$-qubit generalization) differs from
previous methods ($\<1a\>$) in that it only requires Pauli correction.
%
%
This allows the removal of the correction procedure in some of our
schemes as in the $1$-way quantum computer,\cite{Raussendorf01}
providing further hint that the two measurement models are related.
%

Various open questions remain.  We will explore the relation between
the two measurement models in the future.  Universality of
measurements that are not maximally entangling remains to be
investigated.  Finally, error correction, fault-tolerance, and error
thresholds remain to be investigated in detail.

Though experimental advantages of measurement-based quantum
computation are yet to be found, alternative models for quantum
computation and their universality requirements are important for new
experimental directions and insights on what makes quantum computation
powerful.

\section*{Acknowledgments}

We thank Michael Nielsen and David DiVincenzo for interesting
discussions motivating part of the current result.  Jim Harrington
corrected a mistaken omission of $Z$ in $S_{0,1,2,3}$, and Daniel
Gottesman drew our attention to Ref.~\refcite{Gottesmanpriv}.  We
thank Allen Knutson and Eric Rains for insightful ideas on the
universality of the set $\<P, U^\dagger P U\>_{P \in {\cal P}_2}$.
DWL is indebted to Charles Bennett, Isaac Chuang, Beth Ruskai, John
Smolin, and Barbara Terhal for helpful discussions and encouragements.
Duncan Mortimer made an extensive study of measurement models in his
senior thesis, University of Queensland, 2002.
Part of this work was completed when DWL was at IBM TJ Watson Research
Center and at ITP, UCSB.  DWL is partially supported by the US NSF
under grant no. EIA-0086038 and by the Richard C. Tolman Endowment
Fund at Caltech.




\begin{thebibliography}{0}

\bibitem{DiVincenzo95a}
D.~P. DiVincenzo, {\it Science} {\bf 270},  255  (1995). 

\bibitem{Universal} See chapter 4 in \refcite{Nielsen00} for a summary
of the major results and historical notes. A partial list of original
contributions include Refs.
\refcite{DiVincenzo95b,Barenco95a,Boykin99,Lloyd95a,Deutsch95a,Barenco95b}.

\bibitem{DiVincenzo95b}
D.~P. DiVincenzo, {\it Phys. Rev. A} {\bf 51},  1015  (1995).


\bibitem{Barenco95a} A. Barenco, C. Bennett, R. Cleve, D. DiVincenzo,
N. Margolus, P. Shor, T. Sleator , J. Smolin, and H. Weinfurter,
Phys. Rev. A {\bf 52}, 3457 (1995).

\bibitem{Boykin99}
P.~O. Boykin {\it et~al.},  in {\em Proc. 40$^{th}$ Annual Symposium on
  Foundations of Computer Science} (IEEE Computer Society Press, Los Alamitos,
  CA, 1999).  

\bibitem{Lloyd95a}
S. Lloyd, {\it Phys. Rev. Lett.} {\bf 75},  346  (1995).

\bibitem{Deutsch95a}
D. Deutsch, A. Barenco, and A. Ekert, {\it Proc. R. Soc. London A}
{\bf 449}, 669 (1995).  

\bibitem{Barenco95b} A. Barenco, 
quant-ph/9505016.

\bibitem{Shor96} P. Shor, in {\em Proc. 37$^{th}$ Annual Symposium on
Foundations of Computer Science} (IEEE Computer Society Press, Los
Alamitos, CA, 1996), p.\ 56.

\bibitem{Gottesman99}
D. Gottesman and I. Chuang, {\it Nature} {\bf 402},  390  (1999).

\bibitem{Zhou00} X. Zhou, D. Leung, and I. Chuang, {\it Phys. Rev. A'}
{\bf 62}, 052316 (2000).

\bibitem{Bennett93} C.~H.~Bennett, G.~Brassard, C.~Cr$\acute{e}$peau,
R.~Jozsa, A.~Peres, and W.~K.~Wootters, {\it Phys. Rev. Lett.} {\bf
70}, 1895 (1993).

\bibitem{Knill01}
E. Knill, R. Laflamme, and G. Milburn, {\it Nature } {\bf 409},  26  (2001). 

\bibitem{Lidar02}
L.-A. Wu and D. Lidar,  
{\it Phys. Rev. A} {\bf 67}, 050303 (2003). 




\bibitem{Raussendorf01} R. Raussendorf and H.~J. Briegel, {\it
Phys. Rev. Lett.} {\bf 86}, 5188 (2001).

\bibitem{Briegel00} H.~J. Briegel and R. Raussendorf, {\it
Phys. Rev. Lett.} {\bf 86}, 910 (2001). 

\bibitem{Nielsen01p} M.~A. Nielsen, 
quant-ph/0108020v1.

\bibitem{Fenner01}
S.~A. Fenner and Y. Zhang, quant-ph/0111077.

\bibitem{Leung01mit} Leung proved that $3$-qubit measurements are
universal (guest lecture, MIT MAS.961, Oct.~2001) because {\sc cnot}
only has Pauli corrections and its ancilla can be prepared by
measuring $IZZZ$ on ${1 \over 2 \sqrt{2}}
(|00\>+|11\>)(|0\>+|1\>)^{\ot 2}$.  Nielsen subsequently noted that
the ancilla for any $2$-qubit gate has stabilizer generators of weight
$\leq 3$, and can be prepared by $3$-qubit measurements.


\bibitem{Leung01} D. Leung, 
quant-ph/0111122.

\bibitem{Gottesman97}
D. Gottesman, Ph.D. thesis, CalTech, Pasadena, CA, 1997,
 quant-ph/9705052.

\bibitem{Nielsen00}
M.~A. Nielsen and I.~L. Chuang, {\em Quantum computation and quantum
  information} (Cambridge University Press, Cambridge, U.K., 2000).

\bibitem{Leung00}
D.~W. Leung, Ph.D. thesis, Stanford University, Palo Alto, CA, 2000,
cs.CC/0012017.

\bibitem{newtrick}
This trick has been implicitly used by Gottesman~\cite{Gottesman97} to
  construct fault tolerant Clifford group gates for general stabilizer codes.

\bibitem{Gottesman98}
D. Gottesman, 
  quant-ph/9807006.

\bibitem{Gottesmanpriv}
G. Nebe, E. Rains, and N. Sloane, 
math.CO/0001038.  



\end{thebibliography}
\end{document}